\documentclass[11pt,twoside]{article}
\usepackage{asp2010}
\usepackage{subfigure}
\usepackage[top=5.2cm, bottom=2.50cm]{geometry}

\resetcounters

\begin{document}
\label{authorguide}

\title{Broadband nulling behaviour of PSR B2319+60}
\author{V.~Gajjar$^1$, B.~C.~Joshi$^1$, and M.~Kramer$^2$ 
\affil{$^{1}$National Centre for Radio Astrophysics, Post Bag 3, Ganeshkhind, Pune 411 007, India \\
$^{2}$MPI fuer Radioastronomie, Auf dem Huegel 69, 53121 Bonn, Germany}}

\begin{abstract}
Pulse nulling is one of many single pulse phenomena 
exhibited by radio pulsars. The broadband nature of this phenomena remained
unexplained due to lack of coordinated observations. 
We are reporting broadband nulling behaviour of a 
well known nulling pulsar PSR B2319+60.
The simultaneous observations were carried out at
four different frequencies using the Giant
Meterwave Radio telescope (325 and 610 MHz), 
the Westerbork Synthesis Radio Telescope (1420 MHz) 
and the Effelsberg radio telescope (5100 MHz). 
The nulling fractions were estimated 
at all four observed frequencies,
which suggest similar degree of nulling 
across these frequencies. 
To comment on the broadband behaviour of 
pulse nulling, we calculated the Pearson 
cross-correlation coefficients for the  
occurrence of null and burst pulses 
between all four frequencies. 
We conclude that nulling 
is largely a broadband phenomenon for this pulsar 
and it points towards a global failure of the 
magnetospheric currents which produces radio 
emission at these frequencies.
\end{abstract}

\section{Introduction}
Pulsar nulling is one of many single 
pulse phenomena exhibited by radio pulsars. 
A pulsar is said to null when it shows no emission 
for several periods before abruptly resuming its 
normal pulsation. Previous studies tried to relate 
nulling to pulsar ages and the profile morphology, 
but no clear correlation exists. Hence, no conclusive model 
can be given to explain the cause of nulling in pulsars. 

The hypotheses regarding the pulse nulling can be divided 
into two groups. Theories regarding the 
missing sight-lines and/or partial illumination of 
the radio beam \cite[]{Her+07,Her+09} suggest 
geometric effects as the cause of pulsar nulling. 
Alternatively, intrinsic  effects like complete cessation 
of pair production at polar gap, 
loss of coherence and/or break in the 
two stream instabilities can also 
cause the pulsar to null. 
\cite[]{Fil+82,cr80}. 

Pulsar emission at various radio frequencies 
comes from different heights.  Hence, 
the geometry of emitting regions are 
expected to be completely different
for different frequencies.  
The broadband nature of nulling phenomena can shed 
light on the possible hypotheses regarding the 
occurrence of nulling in pulsars. 
If the pulsar emission gets extinguished at 
the polar cap then one would expect its effect at all 
the radio frequencies. Earlier reported broadband nature of 
nulling behaviour suggest that nulling does not always 
occur simultaneously at all frequencies \cite[]{bgk+07,dls+84}. 
However, \cite{bgg08} has reported correlated nulling 
in their multifrequency observations (at 325 and 610 MHz) 
of B0826$-$34. Hence, the question of broadband behaviour of pulse nulling 
remains open due to lack of such coordinated observations. 
More nulling pulsars need to be studied with higher 
sensitivity for this purpose. 

Observations of PSR B2319+60, carried out simultaneous 
at four different frequencies (i.e. 325, 610, 1420 and 5100 MHz) using 
three different telescopes (i.e. Giant Meterwave Radio Telescope; GMRT, 
Westerbork Synthesis Radio Telescope; WSRT and Effelsberg), are 
presented in this paper. 

\section{Observations and Analysis}
The observations of PSR B2319+60 were carried out 
at the GMRT simultaneously at 325 and 610 MHz.
The antennae were divided into two separate sub-arrays, 
one at each frequency. The signal were processed by the GMRT 
software backend, which is a fully software base real-time backend with 
33.33 MHz of instantaneous bandwidth \cite[]{rgp+00}. 
The data were recorded in the total intensity mode 
with nearly 1 millisecond integration time
for both the frequencies. Data were converted to 
SIGPROC\footnote{http://sigproc.sourceforge.net}
filterbank format during the off-line processing. 
At WSRT\footnote{Authors would like to thank Roy Smits for the valuable 
support during the WSRT observations and off-line 
data reduction.},the pulsar was observed at the central 
frequency of 1420 MHz with 160 MHz of total bandwidth using 
standard WSRT Pulsar Machine (PuMa). 
For these observations, we initially recorded 
data with upgraded PuMa II in a raw voltage mode with 
full polarizations at the Nyquist sampling rate. 
In offline processing, data were averaged to 
effective sampling time of around 1 millisecond with total 
intensity before the conversion to SIGPROC filterbank format.
At Effelsberg\footnote{Authors would also like to thank 
Ramesh Karuppusamy, Joris Verbiest and David Champion 
for the valuable support during the Effelsberg 
observations.}, our observations were carried out at 
5100 MHz with total 500 MHz of observing bandwidth. 
We recorded the data in total intensity mode 
with the effective integration time of around 1 millisecond. 
PSRFFTS pulsar search receiver directly generated data 
in the standard SIGPROC filterbank format. 

\begin{figure}[h]
 \centering
 \subfigure[]{ 
 \includegraphics[width=5 cm,height=6 cm,angle=-90,bb=50 50 600 750]{Energyon_comb_forPaper.epsi}
 \label{energyon}
 }
 \subfigure[]{ 
 \includegraphics[width=5 cm,height=6 cm,angle =-90,bb=50 50 600 780]{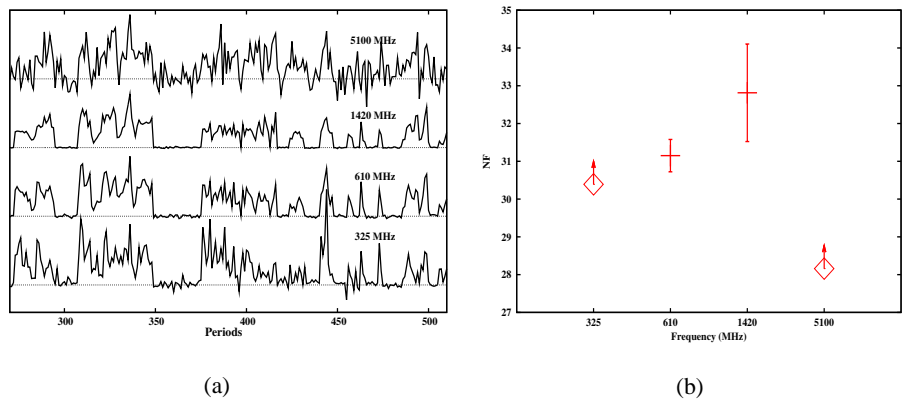}
 \label{NF_vs_freq}
 }
 \caption{(a) Normalized on-pulse energy time series as a function of 
 observed period numbers at four observed frequencies. 
 The on-pulse energy shows similar intensity fluctuations 
 with a hint of correlated nulling behaviour.    
 (b) The estimated NF as a function of observed 
 frequencies. S/N was not sufficient 
 at the observing frequencies of 325 MHz and 5100 MHz, 
 hence only a lower limit on the NF were estimated.}

\end{figure}

The data were dedispersed using the normal 
DM of 94.6 pc$-$cm$^{-3}$\cite[]{hlk+04} for this pulsar 
at all observed frequencies. The dedispersed time 
series were further aligned after correcting the dispersion delay 
between the four observed frequencies. Only the 
aligned time series were used for further 
analysis. The on-pulse energies of these 
aligned time series are shown in Figure \ref{energyon}, 
which clearly shows correlated nulling. 
The nulling fraction (NF) were estimated using the method 
described in \cite{gjk+12} and are shown in Figure \ref{NF_vs_freq}
at all four observed frequencies.  
It was possible to obtain meaningful NF only 
at 610 and 1420 MHz, while lower limits 
were estimated at 325 and 5100 MHz 
using the method described in \cite{gjk+12}. 
The estimated NF and the lower limits 
on the NFs from Figure \ref{NF_vs_freq} 
clearly shows that the degree of nulling 
at four observing frequencies are consistent 
with each other within the error bars.  

The signal to noise ratio (S/N) was sufficient 
to identify single pulses at observing frequencies of 610 and 1420 MHz. 
We identified individual null and burst pulses 
from the observed sequence of single periods. 
To compare the time scale of nulling at these two 
frequencies, the null length histograms (NLHs) 
and the burst length histograms (BLHs) were obtained as described in 
\cite{gjk+12}. The NLH (and BLH) at 610 and 
1420 MHz were compared using the standard 
two sample Kolmogrov-Smirnove (KS) test. 
The null hypotheses assumes 
that both the distributions (i.e. NLH or BLH 
obtained at two frequencies) come from different 
populations. The KS-test comparison with NLH at 1420 MHz 
and 610 MHz rejected the null hypotheses with 95\% significance.
Similarly, the BLH comparison rejected the null hypotheses 
with significance of 97 \%. These tests again demonstrated 
the high degree of correlated nulling between 610 and 1420 MHz. 
It was not possible to build the NLH and BLH at 325 and 5100 MHz 
because of the lower S/N. 

\begin{figure}[h]
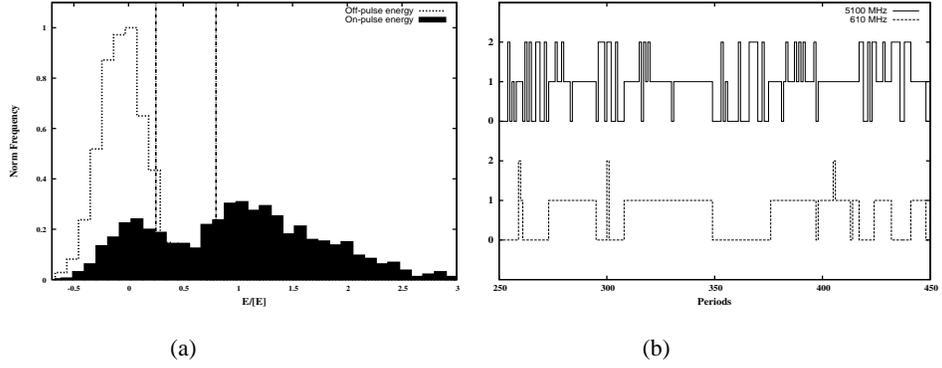

 \centering
 \subfigure[]{
 \includegraphics[width=5 cm,height=6 cm,angle=-90,bb=0 0 600 684]{histogram_for_paper5100.epsi}
 \label{histogram}
 }
 \subfigure[]{
 \includegraphics[width=5 cm,height=6 cm,angle=-90,bb=0 0 600 683]{one_zero_two_paper.epsi}
 \label{one_zero_two}
 }
\caption{(a) On-pulse (filled) and off-pulse (dotted line) 
energy histograms at 5100 MHz. 
The lower and higher thresholds 
(dot-dash-dot line) are shown at 0.3 and 0.7 times the mean pulse energy. 
(b) The zero-one-two time series derived from single pulse data at two of the observing 
frequencies using the procedure described in the text. 
The 5100 MHz time series shows many pulses at level two 
as it had more number of low energy pulses compared 610 MHz. The 
correlated nulling can also be seen between these two series.}
\end{figure}
  
To correlate the occurrence of nulling across all four 
frequencies, we compared individual pulse nulling 
by carrying out a cross-correlation test as follows. 
As mentioned above, the S/N were low for 
many single pulses at 325 and 5100 MHz which 
makes it difficult to separate the low energy 
pulses from the null pulses. To avoid any 
wrong identification, we used two different 
thresholds to separate the null and burst pulses
as shown in the on-pulse histogram of Figure \ref{histogram}. 
Pulses below the lower threshold were consider as 
null pulses and tagged as 0s, while the pulses 
above the higher threshold were confirmed as burst 
pulses and tagged as 1s. The pulses between 
these two threshold were low energy pulses 
and were tagged as 2s. After separating the pulses 
in these three groups, we formed a  
zero-one-two series from the order of their 
observed occurrence which is also shown in 
Figure \ref{one_zero_two}. As these 
time series are aligned in time, 
a cross-correlation test between them 
will check the simultaneity of null and burst 
pulses. We carried out a Pearson correlation 
coefficient comparison test between various 
frequency pairs. Only those pulses were used in 
correlation test which were tagged as either 0s 
or 1s at both the frequencies. Pulses which were 
tagged as 2s were excluded from the correlation test. 
The Pearson correlation coefficients 
between various frequency pairs are shown in Table 1. 
It can be seen from the Table 1 that 
all pairs of frequencies show high value of 
cross-correlation coefficient. This 
suggest highly correlated pulse nulling 
across all observed frequencies. 

\begin{table}
\centering
 \begin{tabular}[h]{|c|c|c|c|}
\hline
Freq(MHz) & 610  & 1420 & 5100 \\
\hline
325   & 0.93 & 0.91 & 0.81 \\
\hline
610   & $-$  & 0.99 & 0.86 \\
\hline
1420  & $-$  & $-$  & 0.85 \\
\hline
 \end{tabular}
\label{table}
\caption{The Person cross-correlation 
coefficients between various observed 
frequency pairs.}
\end{table}

\section{Discussion and conclusions}
We have carried out simultaneous 
multifrequency observations of PSR B2319+60 
at four different frequencies. 
Comparison of NFs and length of nulls 
suggest highly correlated nulling behaviour. 
We have shown that PSR B2319+60 shows high 
degree of correlated nulling across wide 
range of frequencies using the Pearson 
cross-correlation test. Geometric reasons 
are not sufficient to explain the correlated 
nulling across wide range of frequencies. 
Hence, our observations suggests 
the global failure of magnetospheric 
currents as a more likely cause of pulse nulling 
in this pulsar. 

\bibliographystyle{asp2010}
\bibliography{psrrefs.bib,modpsrrefs.bib,mybib.bib}

\begin{thebibliography}{}
\expandafter\ifx\csname natexlab\endcsname\relax\def\natexlab#1{#1}\fi
\expandafter\ifx\csname url\endcsname\relax
  \def\url#1{\texttt{#1}}\fi
\expandafter\ifx\csname urlprefix\endcsname\relax\def\urlprefix{URL }\fi
\providecommand{\eprint}[2][]{\url{#2}}

\bibitem[{{Bhat} et~al.(2007){Bhat}, {Gupta}, {Kramer}, {Karastergiou}, {Lyne},
  \& {Johnston}}]{bgk+07}
{Bhat}, N.~D.~R., {Gupta}, Y., {Kramer}, M., {Karastergiou}, A., {Lyne}, A.~G.,
  \& {Johnston}, S. 2007, A\&A, 462, 257. \eprint{arXiv:astro-ph/0610929}

\bibitem[{{Bhattacharyya} et~al.(2008){Bhattacharyya}, {Gupta}, \&
  {Gil}}]{bgg08}
{Bhattacharyya}, B., {Gupta}, Y., \& {Gil}, J. 2008, MNRAS, 383, 1538.
  \eprint{0711.0526}

\bibitem[{Cheng \& Ruderman(1980)}]{cr80}
Cheng, A.~F., \& Ruderman, M.~A. 1980, ApJ, 235, 576

\bibitem[{Davies et~al.(1984)Davies, Lyne, Smith, Izvekova, Kuzmin, \&
  Shitov}]{dls+84}
Davies, J.~G., Lyne, A.~G., Smith, F.~G., Izvekova, V.~A., Kuzmin, A.~D., \&
  Shitov, Y.~P. 1984, MNRAS, 211, 57

\bibitem[{{Filippenko} \& {Radhakrishnan}(1982)}]{Fil+82}
{Filippenko}, A.~V., \& {Radhakrishnan}, V. 1982, \apj, 263, 828

\bibitem[{{Gajjar} et~al.(2012){Gajjar}, {Joshi}, \& {Kramer}}]{gjk+12}
{Gajjar}, V., {Joshi}, B.~C., \& {Kramer}, M. 2012, \mnras, 424, 1197.
  \eprint{1205.2550}

\bibitem[{{Herfindal} \& {Rankin}(2007)}]{Her+07}
{Herfindal}, J.~L., \& {Rankin}, J.~M. 2007, MNRAS, 380, 430

\bibitem[{{Herfindal} \& {Rankin}(2009)}]{Her+09}
--- 2009, MNRAS, 393, 1391. \eprint{0802.0881}

\bibitem[{Hobbs et~al.(2004)Hobbs, Lyne, Kramer, Martin, \& Jordan}]{hlk+04}
Hobbs, G., Lyne, A.~G., Kramer, M., Martin, C.~E., \& Jordan, C. 2004, MNRAS,
  353, 1311

\bibitem[{{Roy} et~al.(2010){Roy}, {Gupta}, {Pen}, {Peterson}, {Kudale}, \&
  {Kodilkar}}]{rgp+00}
{Roy}, J., {Gupta}, Y., {Pen}, U.-L., {Peterson}, J.~B., {Kudale}, S., \&
  {Kodilkar}, J. 2010, Experimental Astronomy, 28, 25. \eprint{0910.1517}

\end{thebibliography}

\end{document}